\documentclass[11pt,dvips]{article}
\usepackage{epsfig,times}
\usepackage[russian,english]{babel}
\usepackage[cp1251]{inputenc}

\begin{document}
\begin{center}
{\bf \Large Investigation of density structure of pseudorapidity fluctuations in interactions of Au 10.7 AGeV and Pb 158 AGeV nuclei with photoemulsion nuclei by Hurst method}

\vspace{2mm}

{\it \large A.Sh. Gaitinov, P.B. Kharchevnikov, I.A. Lebedev, A.I. Lebedeva}

\vspace{2mm}

{\large Institute of Physics and Technology, Almaty, Kazakhstan}

\vspace{2mm}

\end{center}

{\small 
Analysis of pseudorapidity fluctuations in interactions of 10.7 AGeV $_{79}Au^{197}$ and 158 A GeV $_{82}Pb^{208}$ nuclei with photoemulsion nuclei by Hurst method, is carried out. As result, events of explosive and mixed types, are discovered. These types have different fragmentation characteristics. The most events of the explosive type are processes of full destruction of projectile nucleus, in which multicharge fragments are absent. Events with several multicharge fragments of projectile nucleus are processes of mixed type. Moreover, these two types are separated in multiplicity distribution. Relative number of events of mixed type for 10.7 Au+Em interaction is significantly more than for 158 Pb+Em. Quantity of events of explosive type is approximately equal for both case.}

\large

\section{Introduction}

    Search of phase transition of matter from hadron conditions to quark-gluon plasma is one of the important problems not only the nuclear physics, but also physics in general [1-3]. Experimental and theoretical researches in this area traditionally concentrate on studying of interactions of nuclei at high energies, creating the best conditions for studying of such problems: high pressures and temperature in reaction volume. 

Theoretical estimations show, that in such interactions the energy density can exceed 5 GeV/fm$^3$, i.e. it exceeds the critical density, which is necessary for formation of a new condition of matter, so called the quark-gluon plasma. 

The problems were studied for long time on AGS (Au+Au interactions at energies per nucleon 2.5 - 4.3 GeV), on SPS in CERN (Pb+Pb, 8-150 GeV) and on collider RHIC in Brukheven (Au+Au, 56, 130 and 200 GeV) [4-6]. 

     Qualitatively new results have shown researches of last years. We will underline two basic conclusions of these experiments, - especially RHIC - experiments. The first: in interactions of heavy nuclei there is generation of a quark - gluon plasma with new unexpected properties - properties of an ideal liquid. As result, new scientific representation about strong interacting quark-gluon plasma – sQGP, was generated [7]. The second: mutiparticle processes in the central interactions of heavy nuclei are not represent a superposition of nucleon-nucleon interactions. Scattering of secondary particles from the formed clot of a nuclear matter finds out collective character already at level constituent quarks of these particles, i.e before hadron stages of multiparticle generation [8]. 

     In the latest time, purposeful researches on a problem of an analysis and a search of the mixed phase «excited hadron matters», including free quarks and gluons together with  protons and neutrons, have deduced this actual problem on new level of consideration. In "old" representations it was expected, that phase transition with hadron deconfinment will occur in interaction of nuclei at very high energies. The newest researches have shown, that with the greatest probability the mixed phase of matter is realised at rather low energies, in the range from 4 to 9 GeV per nucleon in system of the centre of mass, at so-called energy of “Dubna glade”. The reached result has caused not only a new sight at a research problem «excited hadron matters», but also has led now to new purposeful concentration both theoretical, and experimental researches in this area [9].
    
Joint institute of nuclear researches - JINR, Dubna - specially for the problem decision of hadron phase transitions prepares new collider experiment NICA. It will propose to reorganize the scheme of carrying out of new experiments on purpose to come nearer to an energy interval of “Dubna glade”. 

In our previous work \cite{leb} we have proposed a normalized range
method for
analysing pseudorapidity correlations in multiparticle production
processes.
This method allows not only to discriminate dynamic correlations from
statistical ones, but also to determine a "force" and "length" of these
correlations.

In the present paper we apply this method to the analysis of experimental pseudorapidity distributions of secondary particles obtained in 10.7 A GeV $ ^{197} Au$ and 158 A GeV $ ^{208}Pb$ interactions with emulsion nuclei.
Main purpose is search of unexpected effects in comparative analysis of interaction of nuclei, which have approximately equal masses, but significantly different energies. At that, one of nuclei is located in region of “Dubna glade”.

\section{Analysis procedure}

As result of high-energy interaction of two nuclei plenty of secondary
particles is produced. According to the existing notions,
secondary particles, which are "emited" from "interaction volume", have
pseudorapidities, corresponding to a central region of pseudorapidity
distribution. At borders of the distribution, fragments (of the target-nucleus
and projectile-nucleus) bring in considerable contribution. And so, in order
to investigate of pseudorapidity
correlations in distribution of particles from "interaction volume"
we have chosen central pseudorapidity interval $\Delta \eta =  4$.
This interval has been subdivided into $k$ parts with $\delta\eta=\Delta\eta /k$.
By counting the number of particles in each subinterval we arrive at a
sequence $n_i^e$.

A pseudorapidity fluctuation, or the normalized relative deviation
of an individual event from average pseudorapidity distribution
\footnote{It is possible to use also absolute deviation
$\xi_i =  n_i^e/n^e - n_i/n.$} is given by
\begin{equation}
\xi_i = \frac{n_i^e/n^e - n_i/n}{n_i/n}= \frac{n_i^e}{n^e} \;
\frac{n}{n_i} - 1
\end{equation}
where $ n_i^e$ is the number of particles in the i-th bin of an event
with
particles number $n^e$, and $n_i= \sum_en_i^e$ is the total number of
particles
for all events in the i-th bin, and $n= \sum_en^e$ is the total number
of
particles for all events.

As described in our previous work \cite{leb} for an investigation of
pseudorapidity correlations we analysed the normalized range
$H(k')= R(k')/S(k')$ (where $R(k')$ and $S(k')$ are a "range" and a standard
deviation, which is calculated by Eqs.(\ref{s})-(\ref{1}), see below)
versus the size of the pseudorapidity interval $d\eta =  k'\delta\eta$,
($1\le k' < k$) using a function
\begin{equation}
H(k')= (a k' )^h
\label{6}
\end{equation}
where $a$ and $h$ are two free parameters and $h$ is the correlation
index (or Hurst index). If the signal $\xi_i$ represents white noise
(a completely uncorrelated signal), then $h= 0.5$.
If $h>0.5$, long-range correlations are in a system \cite{h,f}.

In our calculations we have used $k= 8192$. The choice of such large value
of $k$ is not necessary (it is possible using lesser $k$ also). But
the more $k$ the more accuracy of method is approachable.

So, for the sequence $\xi_i$, $1\le i\le k$,
quantities of $R(k)$ and $S(k)$ were calculated by following formulas:
\begin{equation}
S(k)= \left[ \frac{1}{k} \sum^{k}_{i= 1} [\xi_i-<\xi >]^2\right]^{1/2}
\label{s}
\end{equation}
\begin{equation}
R(k)= \underbrace{max X(m,k)}_{1\le m\le k} -\underbrace{min
X(m,k)}_{1\le m\le k}
\end{equation}
where the quantity $X(m,k)$ characterizes the accumulated deviation from
the average
\begin{equation}
< \xi > = \frac{1}{k} \sum^{k}_{i= 1}\xi_i
\end{equation}
for a certain interval $m\delta\eta$,
\begin{equation}
X(m,k)= \sum^{m}_{i= 1} [\xi_i-<\xi >], \quad 1\le i \le m \le k
\label{1}
\end{equation}

Then the sequence $\xi_i $ has been subdivided into two parts: $\xi_i'$,
$1\le
i\le k'= k/2$ and $\xi_i''$, $k'+1\le i\le k$, and the value of
$H(k/2)= R(k/2)/S(k/2)$ was found for each of the two independent
series.
Similarly $\xi_i'$ and $\xi_i''$ have been subdivided further, followed
by the
calculation of $H(k/4)= R(k/4)/S(k/4)$. This subdivision and analysis
procedure
for newly obtained series of $\xi_i$-values is continued until $d\eta >
0.1$.
$H$ corresponding to the same value of $k'$ have been averaged and drawn
on a
log-log scale as a function of $k'$.

\section {Data}

The above analysis procedure has been applied to experimental data of the $ ^{208}Pb$ 158 A GeV and the $ ^{197}Au$ 10.7 A GeV interactions with emulsion nuclei \cite{au}.

    Nuclear photoemulsion in comparison with other approaches, investigating interactions of nuclei, is the most informative. 
At first, it has high spatial resolution. Secondly the nuclear photoemulsion allows to observe an interaction in $4\pi$ geometry of experiment. The most of other methods have essential dead zones in which secondary particles are not registered.   

    Besides, the method of nuclear emulsion allows to define enough  easily and precisely charges of fragments and gives the chance to registration rather small excited nucleus-targets, and also has no energy threshold of registration of fragments of a nucleus-projectile.

    Moreover, the emulsion technique allows to investigate parameters of fragmentation and collective processes in the same interaction.

\section {Results and discussion}

Detailed analysis of individual events by Hurst method allows to discover different types of processes, which are characterized by defined behaviour of Hurst curve. 

   On the basis of the described procedure we have calculated for individual events the average Hurst index, $h_{av}$, which was defined by the equation (2).

   Further all events have been conditionally divided into two classes: usual events (with stochastic pseudorapidity fluctuations) and correlated events (with essential multiparticle correlation). 
The selection of events was made on the basis of an average Hurst index $h_{av} = 0.62$.
       In our previous work \cite{leb1} it was shown that the criterion $h_{av} = 0.62$ corresponds to process in which all secondary particles were produced from two-particle decays. And so, this criterion conditionally divides all experimental set into processes, in which certain dynamic multiparticle correlations are observed, and on events, in which multiparticle correlations are absent. 

   Therefore, if Hurst index was more than $0.62$, then event was referred into group of correlated events. If $h_{av}$ was less than 0.62, event has been named uncorrelated (poorly correlated).
   
On the basis of the detailed analysis of individual events, four types of processes have been found out.  

    In Fig.1 four individual events of interaction Pb+Em 158 AGeV, which belong to different types, are presented. 

\begin{figure}[tbh]
\begin{center}
\includegraphics*[width=0.8\textwidth,angle=0,clip]{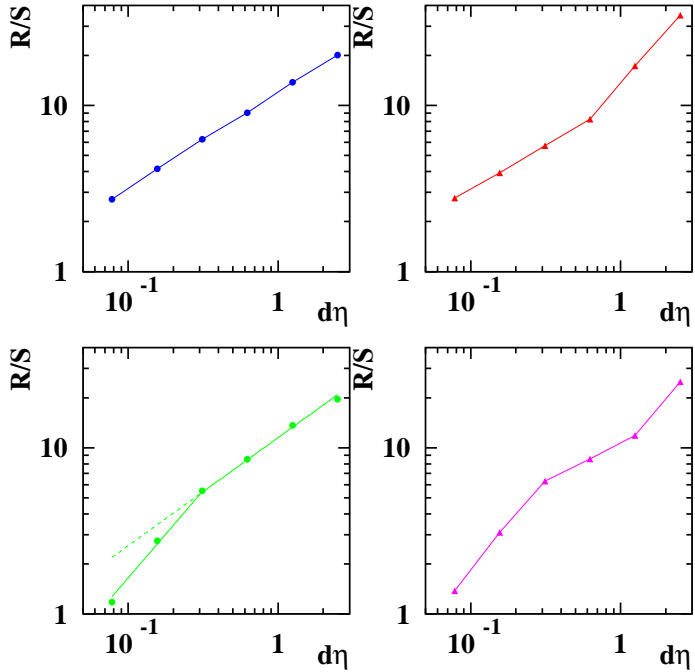}
\caption{\label{fig1} 
Behaviour of Hurst curve for four typical events of Pb+Em 158 AGeV }
\end{center}
\end{figure}

  The first type is characterized by linear behavior of Hurst curve with $h=0.5$ (Fig.1 left-upper). It corresponds to uncorrelated pseudorapidity distribution of secondary particles. Events of such type are well described on the basis of simulating with stochastic pseudorapidity distribution or in the distributions received taking into account two-partial correlations.  

    Events of the second type are characterised by Hurst index $h>0.62$ in the field of small pseudorapidity intervals and $h\sim 0.5$ at other $d\eta$(Fig.1 left-lower). Such behaviour of a correlation curve can be initiated short-range multiparticle correlations. It corresponds to processes of jet type. 

    Events with change of Hurst index to $h>0.62$ in the field of large values of pseudorapidity intervals have been related to the third type (Fig.1 right-upper). Such behaviour of a correlation curve corresponds to essential display of long-range multiparticle correlations and such  events are referred to processes of explosive type.  

    The events with $h>0.62$ both at small values and at large value of pseudorapidity interval, but with $h\sim 0.5$ in middle region of $d\eta$,  have been referred to the fourth type (Fig 1. right-lower). Such behaviour of a curve of Hurst corresponds to events of the mixed type, including processes of explosive type and jet type.
     
To understand mechanisms of formation of final statuses of secondary particles we have analyzed fragmentation parameters of nucleus-nuclear interactions for all types of events. 

As result of the analysis it was discovered that the most events of the explosive type are events of full destruction of projectile nucleus, i.e. event, in which multi-charge fragments are absent. 

Events with several multi-charge fragments of projectile nucleus give Hurst curves corresponding processes of mixed type.

Moreover, these two types are separated in multiplicity distribution.  

    Normalized distributions on multiplicity of events of different  type, are presented in Fig.2.

\begin{figure}[tbh]
\begin{center}
\includegraphics*[width=0.8\textwidth,angle=0,clip]{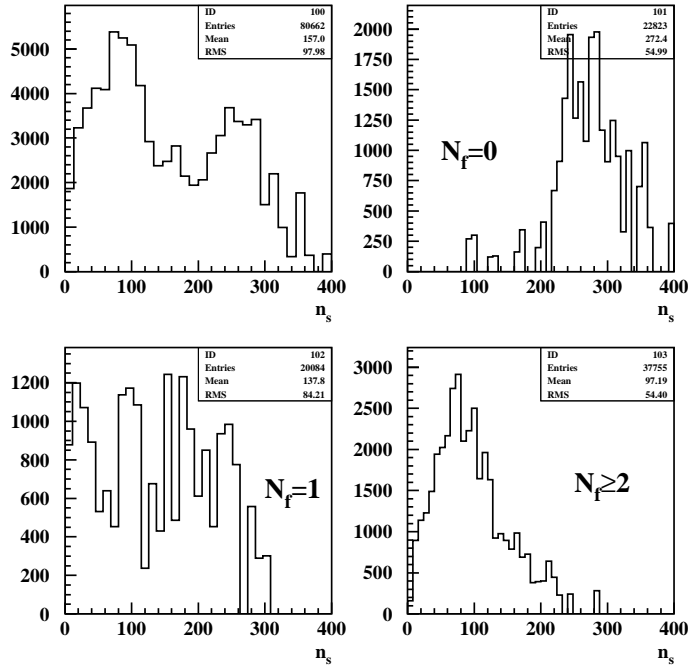}
\caption{\label{fig1} 
ns-distribution, normalized on multiplicity, for events with different number of multi-charge fragments in Au+Em 10.7 AGeV }
\end{center}
\end{figure}
    
As it is seen from the Fig.2, normalized multiplicity distribution is distribution with two clear peaks (Fig.2 left-upper). 

    Moreover, correlation of multiplicity and fragmentation parameters is discovered. 

Events of explosive type, which are characterised by full absence of multi-charge fragments $N_f = 0$ (Fig. 2 right-upper), have large multiplicity. They form the peak with $n_s\sim 280$.

    In events of the mixed type several multi-charge fragments are observed (Fig. 2 right-lower). 

Events with $N_f = 1$ correspond to events of jet type and cascade-evaporative type. The events of these two type have been jointed on the base of the fragmentation parameter. In next analysis these both types were considered in joint group. 

    In table 3.1 the comparative analysis of events of different types in interactions of the heavy nuclei, which have approximately identical mass and charge ($_{82}Pb^{208}$ and $_{79}Au^{197}$), but essentially different energy (158 A GeV and 10.7 A GeV), is presented.    
    
\begin{figure}[htb]
\begin{center}
\tablename{1. Absolute and relative number of events with different types of nucleus-projectile fragmentation in interactions of 10.7 AGeV $_{79}Au^{197}$ and 158 AGeV $_{82}Pb^{208}$ nuclei with photoemulsion nuclei}
\\
\begin{tabular}{|c|c|c|c|c|}
\multicolumn{5}{l}{} \\
\hline
Event type  & \multicolumn{2}{c|}{ Au} & \multicolumn{2}{c|}{ Pb } \\  \hline
$N_f=0$ & \quad 89 events \qquad & 8.1$\%$ & 30 events  & 8.9 $\%$   \\ \hline
$N_f=1$& 406 events & 36.9$\%$ & 189 events  & 52.8 $\%$   \\ \hline
$N_f\ge 2$& 605 events & 55 $\%$ & 137 events  & 38.3 $\%$   \\ \hline
Total& 1100 events &  & 358 events  &    \\ \hline

\end{tabular}
\end{center}
\end{figure}
    
    The analysis of the results, presented in the table, allows to make a conclusion about unexpected dependence of dynamics of multiparticle processes on interaction energy. 

    Following logic assumptions, the increase in energy of interaction of nuclei should lead to increase in events of explosive type which are characterised by full destruction of projectile-nucleus, i.e. full absence of multi-charge fragments $N_f = 0$. Additionally it is expected an increase in events with $N_f\ge 2$.

    However, the relative number of events of full destruction of projectile-nuclei practically has not changed at significant increase of energy ($8.1\%$ of events in interactions Au+Em 10.7 AGeV and $8.9 \%$ in interactions Pb+Em 158 AGeV). 

Moreover, the number of events with $N_f>2$ has essentially decreased ($55 \%$ of events in interactions Au+Em 10.7 AGeV and only $38.3 \%$ in interactions Pb+Em 158 AGeV).  

Thus, given results are unexpected enough and therefore very interesting.

\section {Conclusion}

Investigation of density structure of pseudorapidity fluctuations in interactions of Au 10.7 AGeV and Pb 158 AGeV nuclei with photoemulsion nuclei by Hurst method, is carried out. As result, events of cascade-evaporative, explosive and mixed types, are discovered. 

It is shown that the most events of the explosive type are processes of full destruction of projectile nucleus, in which multi-charge fragments are absent. 

Events with several multi-charge fragments of projectile nucleus give Hurst curves, corresponding processes of mixed type. 

Moreover, these two types are separated in multiplicity distribution.  
Average multiplicity of events of mixed type in interactions Au+Em 10.7 AGeV is $<n_s>\sim 80$.
Events of explosive type form the peak with $<n_s>\sim 270$. 

Events with $N_f = 1$ correspond to events of jet type and cascade-evaporative type.

\vspace{12pt}

{\Large \bf Acknowledgements }

This work was supported by grant N1563/GF of Ministry of Education and Science of Kazakhstan Republic.  




\begin{thebibliography}{99}

\bibitem{1} Dremin I.M.  The quark-gluon medium (micro- and macro- QCD) // Hep-ph, 2011, 1101.5970 v1  
\bibitem{2} E.Shuryak Quark-Gluon Plasma - New Frontiers // J.Phys.G35:104044, 2008
\bibitem{3} H.Hess et al. Heavy-Quark Probes of the Quark-Gluon Plasma at RHIC // Phys.Rev.C73:034913, 2006
\bibitem{4} Schuster T. et al. New results on event-by-event ratio fluctuations in Pb+Pb collisions at CERN SPS energies // nucl-ex 1107.1579 v1; proc. of Quark Matter 2011, the XXII International Conference on Ultrarelativistic Nucleus-Nucleus Collisions
\bibitem{5} Prindle D. et al. Heavy Flavor and Jets at RHIC // nucl-ex 1103.6053 v1; Proc. of the 6th International Conference on Physics and Astrophysics of the Quark Gluon Plasma (ICPAQGP-2010), Goa, India, 6-12 December 2010 
\bibitem{6} K.Wozniak et al. Study of the quark-gluon matter by the PHOBOS experiment // Nucl-ex 2008, 0809.2893 v1 
\bibitem{7} E.Shuryak Physics of Strongly coupled Quark-Gluon Plasma // Prog.Part.Nucl.Phys.62:48-101,2009
\bibitem{8} E.Levin, A.H.Rezaeian Gluon saturation and energy dependence of hadron multiplicity in pp and AA collisions at the LHC // Phys.Rev.D83:114001,2011
\bibitem{9} A.N. Sissakian, A.S. Sorin, V.D. Toneev QCD Matter: A Search for a Mixed Quark-Hadron Phase // Nucl-th, 2006, 0608032, v1, p.23
\bibitem{leb}
I.A.Lebedev, B.G.Shaikhatdenov
J.Phys.G:Nucl.Part.Phys. 23 (1997) 637
\bibitem{h}
H.E. Hurst, R.P. Black, Y.M. Simaika (1965), Long-Term Storage: An
Experemental Study (Constable, London)
\bibitem{f}
J. Feder "Fractals", Plenum Press, New York, 1988,
\bibitem{au}
M.I.Adamovich et al. Phys.Lett. B352 (1995) 1472
\bibitem{10} Bradnova V., Chernyavsky M.M., Gaitinov A.Sh. et all. // Acta physica slovaca. 2004, v.54, No 4, c. 351-365.
\bibitem{leb1} Gaitinov A.Sh., Lebedev I.A., et al. A search of multiparticle correlations in  10.7 A GeV 197Au and 200 A GeV 32S  interactions with emulsion nuclei by the Hurst method // Nucl-th, 2011, 1105.3029 v.1, p.1-10


\end{thebibliography}
\end{document}